\documentstyle[12pt,aps,epsfig]{revtex}
\vspace{2cm}
\begin{document}
\title{\ \\ \ \\ \ \\ \ \\ \ \\  \ \\
From single to many Impurities in one-dimensional systems}
\author{Thierry Giamarchi and H\' el\` ene Maurey}
\address{Laboratoire de Physique des Solides, U.P.S. B\^at 510, 91405
Orsay, France}
\maketitle
\thispagestyle{empty}
\bigskip

{\small We examine the effects of disorder in one-dimensional systems.
We link the case of a few impurities, typical of
a short quantum wire, to that of a finite density of scatterers
more appropriate for a long wire or a macroscopic system.
Finally we investigate the effects
of long-range interactions on the transport in 1D systems. We predict in
that case a conductivity behaving as $\sigma(T) \sim T^2$.}
\bigskip

\section{Introduction}
Since the discovery of Anderson localization, impurity effects
in electronic systems have always been a fascinating subject. When
interactions among the electrons are present, the effect of disorder is
enhanced and leads to various singular behaviors. This is particularly
true in one-dimensional systems, in which interactions themselves have
a dramatic effect on the physical properties of the system and lead to a
non-Fermi liquid behavior.

Prompted by the existing experimental situation of one-dimensional
conductors \cite{jerome_revue_1d},
namely the organic conductors, and by analogy with
the
situation in higher dimensions, studies of such systems mainly looked in
two directions. First the interactions were considered as short-range, a
reasonable assumption for a material with many chains where the long
range Coulomb
interactions could easily be screened. In that case the interactions
lead to the so-called Luttinger liquid behavior
\cite{solyom_revue_1d,emery_revue_1d,haldane_bosonisation}. Second,
a macroscopic system with many weak impurities was considered.
Even in this situation
the disorder has drastic effects and it was rapidly recognized that the
effects of disorder are strongly affected by the interactions
\cite{gorkov_pinning_parquet,mattis_backscattering,luther_conductivite_disorder,%
apel_spinless,apel_impurity_1d,suzumura_scha,giamarchi_loc_lettre,giamarchi_loc}.

More recently progress in nanotechnologies have made it possible to
directly realize a one-dimensional channel of electrons
\cite{thornton_wires,field_1d,scott-thomas_1d,kastner_coulomb,goni_gas1d,%
calleja_gas1d,tarucha_wire_1d,tarucha_quant_cond}.
Edge states of the
fractional quantum Hall effect 
\cite{wen_edge_states,hwang_wires,milliken_edge_states}
 have also been shown to be realizations of Luttinger liquids.
Such systems present a quite different
situation than the organic conductors. First they can be realized of
arbitrary length, and besides the existing disorder
artificial impurities can be made by creating constrictions. For such
systems it is natural to examine the  effects of a few, weak or strong
impurities on the conductance
\cite{glazman_single_impurity,kane_qwires_tunnel_lettre,kane_qwires_tunnel,%
furusaki_1imp,furusaki_2imp,fabrizio_coulomb}

The question that arises is therefore how to link
these two extreme situations, respectively of one or two impurities in a 
system and of a uniform distribution of impurities. We address such a
question in section~\ref{link} and link the methods
used for these two cases.

For realistic wires it is also important to worry about the nature
of the interactions. In a quantum wire, a single channel of electrons
exists, and unless an external gate or metallic plate provides
screening, one can
expect a drastic effect of the long-range Coulomb interactions
\cite{schulz_wigner_1d,glazman_single_impurity}. It is
necessary to reexamine the effects of disorder in the presence of such
long-range forces. In the pure system these forces lead to a dramatic
modification of the physical properties and transform the Luttinger
liquid into the one-dimensional equivalent of a Wigner crystal.
We examine how impurities pin this Wigner crystal in
section~\ref{wigner}, as well as the observable consequences.

\section{Impurities in a Luttinger liquid} \label{link}

Let us consider a one-dimensional interacting system.  We
restrict ourselves in this section to the case of short-range
interactions. The effects of long-range Coulomb interactions will be
examined in section~\ref{wigner}. In that case the system can be
described by the so-called Luttinger Liquid (LL)
\cite{solyom_revue_1d,emery_revue_1d,haldane_bosonisation},
quite different from the normal Fermi liquid occurring in higher
dimensions. To deal with such an interacting system, it is
very convenient to reexpress the excitations of the electron gas in
terms
of collective excitations (charge and spin fluctuations).
This technique, known as bosonization is by now familiar and we will
just recall its salient
features. More details about the technique can be found in
\cite{solyom_revue_1d,emery_revue_1d,haldane_bosonisation}:
for each type of fermions with spin $s=\uparrow,\downarrow$, one introduces 
a field $\phi_s$ related to the
density $\rho_s(x)= -\nabla\phi_s(x)/\pi$. It is convenient to introduce the 
sum
$\phi_\rho$ and the difference $\phi_\sigma$, describing charge and
spin density fluctuations. In terms of such fields the low energy
properties of the system can be expressed for {\bf any} microscopic
Hamiltonian with short-range interactions as $H=H_\rho+H_\sigma$ with
\begin{equation}  \label{quadra}
H_\nu = \frac1{2\pi}\int dx \left[ (u_\nu K_\nu)(\pi \Pi_\nu)^2
         + (\frac{u_\nu}{K_\nu}) (\partial_x \phi_\nu)^2 \right]
\end{equation}
where $\Pi_\nu$ is the momentum conjugate to $\phi_\nu$. {\bf All}
interaction effects are absorbed in the parameters $u_\nu$
(the velocity of charge or spin excitations) and
$K_\nu$. $K_\rho=1$ in the absence of interactions,
$K_\rho > 1$ for attractive interactions and $K_\rho < 1$ for
repulsive ones. In fact the
spin Hamiltonian is slightly more complicated and contains, in addition
to the quadratic part, a sine-Gordon term
\begin{equation} \label{sine}
\frac{2g}{{(2\pi\alpha)}^2}\int dx\, \cos({\sqrt{8}\phi_\sigma})
\end{equation}
which describes the backscattering of electrons with opposite spins.
For spin isotropic interactions $K_\sigma - 1 \sim g/(\pi u_\sigma)$ and
(\ref{sine}) renormalizes the spin part of (\ref{quadra}) to impose
asymptotically $K_\sigma^* = 1$ and $g^*=0$ (for repulsive
interactions). For attractive interactions a gap opens in the spin
excitations and the spin degrees of freedom can be discarded.

The main physical manifestation of the Luttinger liquid is the
nonuniversal decay of the correlation functions. For example the
density-density correlation behaves as
\begin{equation} \label{density}
\langle \rho(x)\rho(0)\rangle=\frac{K_{\rho}}{(\pi x)^2} +
A_1 \cos(2k_Fx)x^{-1-K_{\rho}}
\ln^{-3/2}(x) + A_2 \cos(4k_Fx)x^{-4K_{\rho}}+...
\end{equation}
For the case of spin-anisotropy the exponent of the $2k_F$ part is given
by $-(K_\rho+K_\sigma)$, and in (\ref{density}) the log corrections arise
from the renormalization of $K_\sigma$ to one \cite{giamarchi_logs}.
For not too repulsive interactions the $2 k_F$
fluctuation is the strongest. The $4 k_F$ term comes from the
interplay of the density and the backscattering term (\ref{sine}) and
has
extremely important effects for Coulomb interactions, as will be seen in
section~\ref{wigner}. In what follows we will note by $-\mu$ the
exponent of the density-density correlation function (for a Luttinger
liquid with moderately repulsive interactions $\mu=K_\rho+K_\sigma$).

This nonuniversal decay of the density correlations greatly affects the
coupling to disorder.
Disorder can be introduced in the system by putting impurities at
positions $R_i$. If for simplicity we take an identical potential $V$
for each impurity, the disorder term becomes
\begin{equation}  \label{diso}
H_{\text{dis}} = \sum_{i} \int dr V(r-R_i) \rho(r)
\end{equation}
For macroscopic systems, the impurities are randomly distributed, and
transport is best described by the conductivity, as a
function of either temperature or frequency. In the case of a short
system, or for a few impurities, it is more convenient to compute the
conductance.

\subsection{Many impurities}

Let us first examine the macroscopic system with many random impurities.
For a macroscopic system, and for not too strong disorder,
one usually approximates the disorder by a
random potential $H_{\text{dis}} = \int dr V(r) \rho(r)$
where $V(x)$ is Gaussian correlated $\overline{V(x)V(x')} =
D_\xi\delta(x-x')$. Although the main motivation in doing so is probably
theoretical simplicity, such a substitution is quite good for weak
disorder. Indeed it assumes that the concentration of
impurities becomes
infinite $n_i \to \infty$, but that the potential of each impurity
becomes weak $V \to 0$, so that the product $n_i V^2 = D$,
remains a constant. In the process one loses a
parameter: the average distance between impurities (or the strength of
the potential on {\bf one} impurity).
Such an approximation is in fact equivalent to looking only at
the lowest order in the self energy of the electrons \cite{abrikosov_ryzhkin}.

Various methods have been proposed to treat such a disorder. We will
explain here the results obtained by a renormalization group method
since it allows both to treat {\bf finite} disorder and to make
connection with the single impurity case (for more references on the
other
methods see e.g. \cite{giamarchi_loc}). To use the RG one expands in
powers of the coupling to disorder $D_\xi$. The perturbation is
divergent, and one can derive \cite{giamarchi_loc_lettre,giamarchi_loc}
renormalization equations upon
rescaling of the short distance cutoff $\alpha\to \alpha e^l$
\begin{eqnarray}
\frac{d K_\rho}{d l} = -  \frac{2u_\rho}{u_\sigma} D \label{lesk} \\
\frac{d D}{d l} = D (3 - \mu - g)  \label{lesd}
\end{eqnarray}
where $\mu$ is the exponent of the density-density correlation function
and $g$ is defined in (\ref{sine}).
$D$ is a dimensionless parameter related to the disorder
$D = \frac{2D_\xi\alpha}{\pi u_\sigma^2}
\left(\frac{u_\sigma}{u_\rho}\right)^{K_\rho}$
There are similar renormalization equations for the spin part
but since we focus on transport we will ignore them for the moment.
The $(3-\mu)$ term in equation (\ref{lesd}) for the disorder is easy
to understand and comes from the dimension of the second order term in
disorder
\begin{equation}
\int dx d\tau \int dx' d\tau' D_\xi \delta(x-x') \langle \rho(x,\tau)
\rho(x',\tau') \rangle
\end{equation}
Note that the backscattering term $g$ changes
this factor into $(3-\mu-g)$. Very often this term $g$ is
incorrectly neglected. Two reasons for this: (i) it is difficult to
obtain since it results from
contractions in higher order in the RG, although it is in fact of the
same order as the terms giving $\mu$. No naive derivation of the RG
equations will give it. (ii) For repulsive
interactions $g$ scales to zero and $K_\sigma\to 1$, and it is
argued that to get the correct asymptotic physics one can merely insert
the renormalized values of the parameters in (\ref{lesd}). This is
perfectly correct to obtain the phase diagram, but not if one wants
to describe the physics at {\bf intermediate} length scales (such as
conductivity at finite length or finite temperatures). This term
traduces the fact that a local repulsion among opposite spins, although
unable to block density fluctuations at large length scales (namely
$K_\sigma^* \to 1$) will kill short distance density fluctuations,
making the system
much harder to pin on disorder. At intermediate length scales, repulsive
interactions tend to make the disorder less relevant and decrease
localization (let us again emphasize that this effect is only at
intermediate length scales). This has dramatic
consequences on the effects of interactions on persistent currents
\cite{giamarchi_shastry_persistent}. We will not explore these effects
further (more details and consequences can be found in
\cite{giamarchi_loc,giamarchi_shastry_persistent}) and denote by
$\tilde{\mu} = \mu+g$, the correct factor in (\ref{lesd}).

\begin{figure}
   \centerline{\epsfig{file=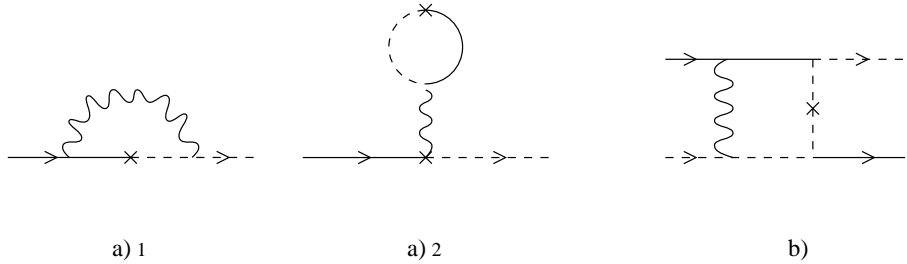,angle=-90,width=12cm}}
   \vspace{0.5cm}
   \caption{Diagrams describing the renormalization of the disorder
by the interactions (a) and the renormalization of the interactions by
the disorder (b). Solid and dotted lines are fermions with $\pm k_F$,
the wiggly line is the interaction and the cross is the impurity
scattering.}
      \label{figure1}
\end{figure}

From equation (\ref{lesd}) is is easy to see that the disorder is
relevant when $\mu < 3$. For repulsive interactions between spins
($g >0$) this
corresponds to $K_\rho < 2$, and for attractive interactions between spins to
$K_\rho < 3$.
(Let us recall that $K_\rho=2$ or $3$ correspond to extremely attractive 
interactions among charges.)
\cite{apel_impurity_1d,suzumura_scha,giamarchi_loc_lettre,giamarchi_loc}.
The equation (\ref{lesd}) traduces the renormalization of the Born
amplitude
of disorder by the interactions and has been derived under various forms in the past
\cite{gorkov_pinning_parquet,mattis_backscattering,luther_conductivite_disorder}.
In a diagrammatic language it corresponds
\cite{gorkov_pinning_parquet,giamarchi_thesis} to diagram
(a) of Figure~\ref{figure1}.

In fact, in the RG treatment the equation (\ref{lesd}) is not sufficient
to correctly describe the physics of the problem as we will see below,
and should be complemented by equation (\ref{lesk}). This first equation
(\ref{lesk}) traduces the effect of disorder on the
exponents of the correlation functions. It has also a  diagrammatic
representation given in Figure~\ref{figure1} (b)
\cite{giamarchi_thesis}.

Using the RG one can compute the finite temperature (or finite
frequency) conductivity of the system
\cite{giamarchi_loc_lettre,giamarchi_loc}. The idea is simply to
renormalize until the cutoff is of the order of the thermal length $l_T
\sim u/T$ corresponding to $e^{l^*} \sim l_T/\alpha$.
At this length scale the disorder can be treated in the Born
approximation. As the conductivity is a physical quantity it is not
changed under renormalization and we have:
\begin{eqnarray} \label{renodens}
\sigma(n(0),D(0),0) = \sigma(n(l),D(l),l) = \sigma_0
\frac{n(l)D(0)}{n(0)D(l)} = \sigma_0 \frac{e^l D(0)}{D(l)}
\end{eqnarray}
where $\sigma(n(l),D(l),l)=\sigma(l)$ and $n(l)$ are respectively
the conductivity and the electronic density at the scale $l$.
$\sigma_0=e^2 v_F^2/2\pi\hbar D_\xi$
is the conductivity in the Born approximation,
expressed with the initial parameters. To compute the conductivity
the full integration of the two equations
(\ref{lesk}-\ref{lesd}) is required
\cite{giamarchi_loc_lettre,giamarchi_loc}.
A simple expression can be obtained for infinitesimal disorder
$D\to 0$  since one can neglect the
renormalization of the exponents (\ref{lesk}). In that case one can
trivially integrate (\ref{lesd}) to obtain, using (\ref{renodens})
\begin{equation} \label{simple}
\sigma(T) \sim \frac1{n_i V^2} T^{2-\tilde\mu}
\end{equation}
This result is quite ancient
\cite{gorkov_pinning_parquet,mattis_backscattering,luther_conductivite_disorder}
and corresponds simply to the
renormalization of the effective disorder by the interactions. One
immediately sees that (\ref{simple}) alone would lead to a paradox since
(\ref{lesd})  gives a localized-delocalized boundary at $\mu=3$
whereas (\ref{simple}) gives perfect conductivity above $\mu=2$ (i.e.
the noninteracting point). This apparent contradiction is solved when
(\ref{lesk}) is correctly taken into account. For
$\mu < 3$ (including the noninteracting point) any small but {\bf
finite} disorder {\bf grows}, renormalizing the exponents
and ultimately leading to a decrease of the conductivity,
even if one started initially from $\mu > 2$. A crude way of taking into
account both equations (\ref{lesk}-\ref{lesd}) would be to say that one
can still use (\ref{simple}) but with scale dependent exponents (see
\cite{giamarchi_loc_lettre,giamarchi_loc})
\begin{equation} \label{complique}
\sigma(T) \sim T^{2-\mu(T)}
\end{equation}
This renormalization of exponents and the faster decay
of conductivity is in fact the
signature of Anderson localization.
The whole RG scheme breaks down when $D\sim 1$, at a length scale
corresponding to the localization length of the system
\cite{giamarchi_loc_lettre,giamarchi_loc}.  A reasonable guess of the
temperature dependence below this length scale is an exponentially
activated conductivity.

\subsection{Single impurity}

Let us now examine the case of a single impurity.
The coupling to disorder is simply
\begin{equation}
H =  V  \rho(x=0)
\end{equation}
and since scattering occurs only in a finite (here one point) portion of
the sample, the conductance is the more appropriate way to describe
transport.
For weak $V$ one can use the same renormalization method expanding
in the interaction $V$. One obtains the RG equations
\cite{kane_qwires_tunnel_lettre,kane_qwires_tunnel}
\begin{eqnarray}
\frac{d K_\rho}{d l} = 0 \label{srg} \\
\frac{d V^2}{d l} = V^2(2 - \tilde\mu)  \label{lesds}
\end{eqnarray}
The first difference in (\ref{srg}) compared to the finite density of
impurities is the absence of renormalization of
the exponents. The second equation (\ref{lesds}) is seemingly different
from the one for Gaussian disorder (\ref{lesd}): The factor $2$, instead
of $3$, now comes from the fact that the impurity only acts at $x=0$ leaving
only a double integral over time.
In fact this difference is only apparent and (\ref{lesds}) and
(\ref{lesd}) are in fact the {\bf same} equation
giving the renormalization of the Born
amplitude of disorder by the interactions. The difference in dimension
can be accounted for by the fact that $D$ is not just the impurity potential
but $D_\xi = n_i V^2$ where $n_i$ is the
impurity concentration, and contains also the inverse of a length. One
can define $\tilde D = e^l D$. Upon renormalization $\tilde D$ follows
the RG equation (\ref{lesds}), but of course would lead to a modified
equation (\ref{lesk}), preserving the physics for the case of a uniform
disorder.

For a single impurity the only effect of interactions is therefore to
change the Born amplitude of scattering (for weak disorder). The
conductance is therefore given by the effective scattering at the scale
$l=\ln(E_F/T)$. Integrating (\ref{lesds}) one gets
\cite{kane_qwires_tunnel_lettre,kane_qwires_tunnel}
\begin{equation} \label{corcond}
G_0 - G(T) = -\delta G(T) \propto \delta R(T) = V^2 T^{\tilde\mu -2}
\end{equation}
where $\delta R$ is the scattering produced by one impurity and $G_0$
the conductance of a pure wire. Of course here since only
renormalization of the disorder is present the transition between
zero/infinite conductivity (equivalently zero/finite conductance) occurs
at $\mu=2$, (i.e. in the vicinity of the noninteracting point).

When the disorder is relevant $\mu<2$, the weak coupling RG scheme
ceases
to be valid when the Born amplitude is of order one. Contrarily to
Gaussian disorder where this indicates the localization, here one has a
different strong coupling fixed point. The pinning on the impurity
becomes strong and one has to consider weak tunneling through the
impurity site. This fixed point has been analyzed in
\cite{kane_qwires_tunnel_lettre,kane_qwires_tunnel} and still gives a
power law for the conductance, but with a different exponent than in the
weak coupling case.

\subsection{Impurities vs Impurity}

A summary of the various equations and physical behaviors for a single
and many impurities is given in table~\ref{table1}.
\begin{figure}
\begin{center}
\begin{tabular}{||c|c|c|c|c|c|c||}
\tableline
disorder & D & $dK_\rho/dl$ & $dD/dl$ & transition & $\sigma$ or
$\delta G$
& strong coupling  \\
\tableline
Gaussian & $n_i V^2$ & $-D(l)$ & $3-\mu(l)$ & strong attraction &
$\sigma \sim T^{2-\mu[T]}$ & Anderson loc. ($\sigma \sim e^{-T^*/T} {\rm
?}$) \\
\tableline
single & $ V^2$ & $0$ & $2-\mu(l)$ & non-interacting &
$\delta G \sim T^{\mu-2}$ & strong barrier ($G\sim T^{1/K_\rho - 1}$) \\
\tableline
\end{tabular}
\end{center}
\caption{Renormalization of the various parameters, physical properties
and transport properties both in the case of
uniform disorder and of a single impurity. For the relation between
$\sigma$ and $\delta G$ see equation (\protect{\ref{gsig}}).}
\label{table1}
\end{figure}

As is obvious from Table~\ref{table1} the two systems offer striking
similarities as well as some important physical differences.
As can be expected on
physical grounds since $D$ has the meaning of $n_i V^2$  at weak
coupling the behavior of a single impurity
corresponds exactly to taking {\bf first} the limit of infinitesimal
disorder $D \to 0$ in the equations (\ref{lesk}-\ref{lesd}) (and
therefore formally to the $n_i\to 0$ limit i.e. a single impurity). This
limit is non trivial since in principle the Gaussian approximation
corresponds to infinitely dense (very weak) impurities
$n_i \to \infty$. However if one takes $D\to 0$ first, since the
equation (\ref{lesd}) does not depend on the strength of $D$, the only
effect is to neglect the renormalization
of the exponents by the disorder. To match the
conductivity and conductance one notices that
in the limit $n_i \to 0$ the impurities are well separated enough to be
considered as independent scatterers and their scatterings add up. For
$N_i$ impurities in
a wire of length $L$ then the conductivity would be, if $G$ is the
conductance of the wire and using $G = \sigma(\omega=0) L$
\begin{equation} \label{gsig}
\sigma(T) = L G \propto \frac{L}{N_i} (\delta R)^{-1} = \frac1{n_i}
(-\delta G)^{-1}
\end{equation}
showing again that (\ref{corcond}) and the approximation (\ref{simple})
are identical.
As we already pointed out, the collective effects of
impurities
(i.e. the fact that they cannot be considered as independent scatterers)
manifest themselves in the renormalization of the exponents. This leads
to a {\bf faster} decay of the conductivity/conductance when
the temperature decreases compared to that of independent impurities.

For Gaussian disorder collective effects manifest themselves
at any length scale since again formally $n_i\to \infty$. On the other
hand the scattering on a single impurity can never become strong.
For realistic (Poissonian) disorder corresponding to many impurities
with both an average distance $a$
between the impurities and a strength of the potential of one impurity
$V$, one can therefore expect a competition between two length
scales: the temperature $T_2$ for which the thermal length $u/T$ is
equal to the distance between impurities and the crossover temperature
$T_{\rm 1,cr}$ for which {\bf one} impurity goes to strong coupling.

As long as the thermal length $u/T$ remains smaller than the
distance between impurities $a$, one cannot have any collective effect
\cite{furusaki_crossover,maurey_qwire}.
There is no renormalization of the exponents, even for many impurities
and the system is described by the equations (\ref{srg}-\ref{lesds}).
The conductivity of the system is given by (\ref{simple}) i.e. with
{\bf fixed} exponents. Two cases are  then possible:

(i) Either the
collective effects between impurities will become important before each
impurity can reach strong coupling ($T_2 > T_{1,cr}$). This is the case
if the impurities
are weak and/or dense enough. In that case below $T_2$ one can use the
Gaussian representation of the disorder (\ref{lesk}-\ref{lesd}), and the
exponents  start to be renormalized. The conductivity becomes
(\ref{complique}) i.e. the same type of temperature dependence,  with
varying exponents. Finally at the scale $T_{\rm loc}$ the disorder
becomes of order one and the system becomes localized.
Below $T_{\rm loc}$ we
expect an activated conductivity, but up to our knowledge no theoretical
method allows to investigate this regime unambiguously. Experimental
measurements of the conductivity in this regime would therefore prove
to be extremely useful. Note that already at the scale $T_{\rm loc}$
the conductivity is a strongly decreasing function of the temperature
since $\mu \sim 0$ and therefore $\sigma(T) \sim T^2$)
\cite{difference_crossover}.

(ii) Impurities are dilute enough or strong enough to reach individually
strong coupling before the collective effects can take place. One then
crosses over to the tunnelling behavior
\cite{kane_qwires_tunnel_lettre,kane_qwires_tunnel}. Collective effects
will occur at a lower temperature but since each impurity corresponds to
a strong potential, they will have to be
examined by quite different methods, leading to a very different
localization transition.
This fascinating problem is quite complicated and has only been solved
up to now for two barriers \cite{kane_qwires_tunnel,furusaki_2imp} but a
solution for many impurities is still lacking.

\begin{figure}
   \centerline{\epsfig{file=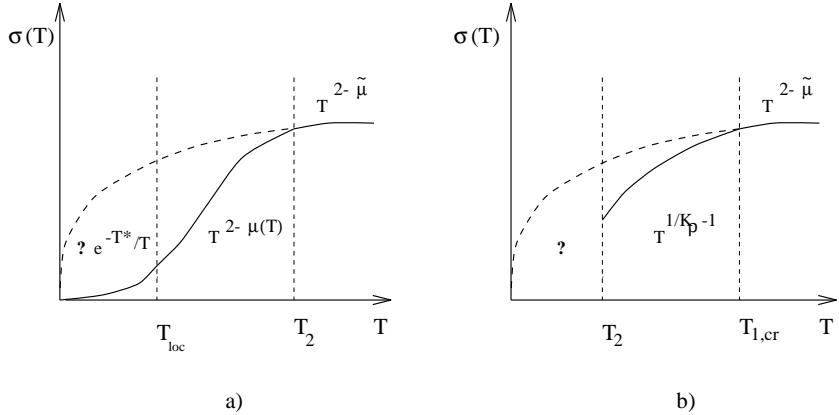,angle=-90,width=11cm}}
   \vspace{0.5cm}
   \caption{Temperature dependence of the conductivity. The dashed line
   is the prolongation of the $T^{2-\tilde{\mu}}$ law. (a) ``Weak
pinning'': $T_{1,cr}<T_2$. Collective effects occur before each impurity can reach
strong coupling. Below $T_2$, renormalization of the exponents gives
a faster decay of the conductivity.
(b) ``Strong pinning'': the opposite situation, $T_{1,cr}>T_2$.}
      \label{figure3}
\end{figure}

These two situations are reminiscent of the
``weak pinning'' and ``strong pinning'' cases of the classical charge
density waves, and could in principle be observed in quantum wires.
The various situations are summed up on figure~\ref{figure3}.

\section{Long-range Coulomb interactions} \label{wigner}

Let us now focus on the effects of long-range interactions on a
one-dimensional system. For quantum wires, it is a priori important to
retain the long-range nature of the Coulomb interaction. Whether this is
screened or not in a given experimental system is an
heavily debated question. At least for some experimental situations
there is evidence that Coulomb interactions play an important role
\cite{goni_gas1d}. We will show here how  transport properties of the
system allow to answer this important question and show characteristic
features of the Coulomb interaction.

It is by now well known that in the presence
of long-range forces the physical properties of the system are quite
different from those of a LL. As we saw from (\ref{density})
density-density fluctuations
have power law decay, and for not too strong repulsion (i.e.
$K_{\rho}>1/3$), dominant density fluctuations are the $2 k_F$ ones.
When long-range interactions are present, the electrons
form a Wigner ``crystal'' (WC)
\cite{schulz_wigner_1d,glazman_single_impurity}: indeed the more slowly
decaying correlation functions are now the $4 k_F$ ones
corresponding to the
distance between electrons. The decay is slower than any power law,
of the form \cite{schulz_wigner_1d}
\begin{equation}
\langle \rho_{4k_f}(x)\rho_{4k_f}(0)\rangle \sim e^{-\ln^{1/2}(x)}
\end{equation}
The $2 k_F$ charge and spin correlations still decay as power laws.
Due to the change in nature of the dominant correlations, one can expect
quite different transport properties than in a LL
\cite{apel_impurity_1d,suzumura_scha,giamarchi_loc_lettre,giamarchi_loc}.
Pinning on a single impurity \cite{fabrizio_coulomb,furusaki_coulomb}
or two impurities \cite{maurey_lettre_2barr,maurey_moriond} already
shows these differences and lead to extremely interesting new behaviors
for the conductance. Here we confine ourselves to the study of
pinning on many weak impurities, such a study being relevant for long
wires.

The WC and the pinning on disorder can be described using
again a bosonization representation similar to (\ref{quadra}).
We will not dwell here on the technical details, that can be found in
\cite{maurey_qwire}, and concentrate on the results.
In the following we will drop the subscript $\rho$ in $K_{\rho}$ and
$\Phi_{\rho}$. The WC can be
described as a modulation of
the charge density $\rho(x) \sim \rho_0 \cos(Qx+2\sqrt{2}\Phi)$ where
$\rho_0$ is the uniform amplitude of the charge density, $Q= 4 k_F$ its
wave vector. $\Phi$ describes here the location and the motion of the
Wigner crystal. Coupling to disorder is
described again by a random potential of the form (\ref{diso}).
Since in the WC the $4 k_F$ density fluctuation is the dominant one
(i.e. the one with the slowest decay), transport will be dominated by
$4 k_F$ scattering on impurities, contrarily to the case of
a LL where $2 k_F$ scattering is usually the dominant one.
This has important consequences for electrons with spins: even in the
presence of long-range charge interaction or very strongly repulsive
short-range interactions $K_\rho\to 0$ spin isotropy still imposes
$K_\sigma^* = 1$, and the $2 k_F$ density-density correlation still
decays at best as $1/r$. Considering the $2 k_F$ scattering on
impurities
as was done in \cite{ogata_wires_2kf,fukuyama_wires_2kf} amounts to
underestimating seriously the scattering on disorder and gives incorrect
exponents for the temperature dependence of the conductivity.
Here we  retain the dominant $4 k_F$ scattering only
\cite{maurey_qwire}.

In the presence of impurities
the Wigner crystal is pinned: the phase $\Phi(x)$ adjusts to the
impurity potential on a scale given by $L_0$ the pinning length (which
corresponds to the localization length of the electron system).
This process of pinning is analogous to what happens in charge density
waves (CDW) \cite{lee_rice_cdw,fukuyama_pinning}, but with important
differences: (i) since we are dealing with electrons,
quantum effects are a priori important contrary
to what happened for charge density waves; (ii) the long-range Coulomb
interactions have to be taken into account.
To study such effects one uses techniques similar
to \cite{lee_rice_cdw,fukuyama_pinning} suitably modified to take into
account (i) and (ii) \cite{maurey_qwire}.
The pinning length is given by
\begin{equation} \label{length}
L_0 =\Biggl(\frac{8e^2/\kappa}{\alpha \pi ^2 V_0\rho _0\gamma
n_i^{\frac1{2}}}\Biggr)
^{\frac2{3}}
\ln ^{\frac2{3}} \Biggl(\frac1{d}
 \Bigl(\frac{8e^2/\kappa}{\alpha\pi^2 V_0\rho _0\gamma
n_i^{\frac1{2}}}\Bigr)
^{\frac2{3}}\Biggr)
\end{equation}
where $\gamma = e^{-4\langle\hat{\Phi}^2\rangle} \approx e^{-\frac{8\tilde{K}}
{\sqrt{3}}\ln^{1/2}V_0}$ and
$\tilde{K} = \frac{\sqrt{\pi uK \kappa}}{2\sqrt{2}e}$. $V_0$ is the
strength of
the impurity potential, $n_i$ their concentration, $d$ the width of the
wire, $\kappa = 4 \pi \epsilon$  the dielectric constant and $K$ the LL
parameter, taking into account the
short-range part of the Coulomb interaction, a number typically of order
$0.5-1$. In the above expression
we have neglected $\log(\log)$ corrections, and, estimating
numerically the relative contributions of the elastic (short-range part
of the interaction $q \sim 2 k_F$) and Coulomb terms  (the long-range
part $q \sim 0$) using
typical values $u=3\times 10^7cm.s^{-1}$ and $K=0.5$), we have kept only
the dominant Coulomb term. For comparison the localization length is
$L_0 \approx (\frac {v_F}{\alpha \pi V_0\rho_0n_i^{1/2}})^{2/3}$ for a
charge density wave and
$L_0 \approx (\frac {v_F}{\alpha \pi
V_0\rho_0n_i^{1/2}})^{2/(3-K_\rho-K_\sigma)}$ for a LL to the same
degree of approximation (for the LL, due to the strong quantum
fluctuations, this expression
is only valid for very small disorder and far from the transition. For a
more complete formula see \cite{giamarchi_loc_lettre,giamarchi_loc}).

Coulomb interactions have two effects. First they give the logarithmic
factor enhancing the rigidity of the system. Secondly they kill the
anomalous exponents coming from the quantum fluctuations (the
$K_\rho+K_\sigma$ in the LL) and drive the system to a classical
limit. This can be traced back to the fact that the correlation
functions decay much more slowly than in a LL ($e^{-\ln^{1/2}(r)}$
instead of a power law), therefore the system is much more ordered and
the fluctuations around the ground state are much less important.
Although they do not lead any more to anomalous exponents, quantum
effects are still important: they strongly reduce  the effective
disorder seen by
the WC since $V \to V \gamma$. This effect can be quantitatively very
important (since $L_0$ is very large for
dilute impurities), and contributes to making the system more likely to
be in the weak pinning regime.
Let us emphasize again that this limit cannot be reproduced naively by
taking the LL and just letting the {\bf charge} interactions becoming
large $K_\rho\to 0$, due to the $K_\sigma$ term. It is crucial to
consider the $4 k_F$ scattering for which spin fluctuations are
absent.

Another important length scale comes from the competition between the
short-range and  long-range parts of the Coulomb interaction.
In addition to the long-range part responsible
for the formation of the WC at large distances, the short distance
repulsion also gives rise to LL effects. Below a certain length scale
$L_{\rm cr}$, the
short-range part is dominant, and the system can be described  by $2
k_F$ scattering on impurities, power law type correlation functions
and standard LL transport. Above this length scale the effect of
the long-range part of the Coulomb interactions is dominant and one
recovers the WC behavior. For a wire with a single mode and no
external screening, we estimated for reasonable parameters
$L_{\rm cr} \sim d$ where $d$ is the width of the wire and for practical
purposes the whole behavior should be WC. For wires where the screening
of the Coulomb interactions is more efficient, one could observe a
crossover between LL and WC behavior.

\begin{figure}
  \centerline{\epsfig{file=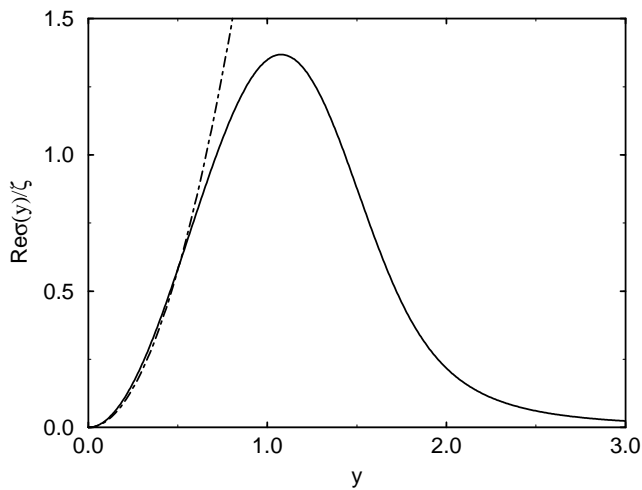,angle=-90,width=6cm}}
   \vspace{0.5cm}
      \caption{Frequency dependence of the conductivity. For simplicity
the regime above $\omega_{\rm cr}$ is not shown. $y$ stands for the rescaled
frequency $\omega/\omega_{\rm pin}$. $\zeta$ is a constant depending on the 
parameters of the wire.
The dash-dotted curve shows an $y^2$ law.}
      \label{freq}
\end{figure}

The most interesting quantity to measure is of course the conductivity
(or conductance). The above two length scales can of course be converted
into either frequency or temperature by using the dispersion relation
$\omega_L = \epsilon(q=1/L)$, and define a pinning frequency
$\omega_{\rm pin}$ (or temperature $T_{\rm pin}$) and a crossover
frequency
$\omega_{\rm cr}$ (or temperature $T_{\rm cr}$). The frequency
dependence of the real part of the conductivity is shown on
figure~\ref{freq}

At small frequencies $\sigma(\omega) \sim \omega^2$, a similar result
than for the LL
\cite{berezinskii_conductivity_log,vinokur_cdw_exact,%
giamarchi_columnar_variat} (up to log corrections).
The low frequency conductivity
is to be contrasted with the previous result of Shklovskii and Efros
\cite{shklovskii_conductivity_coulomb} who find $\sigma(\omega) \sim
|\omega|$.
They derived this result in a very different physical limit when the
localization length is much smaller than the interparticle distance.
In that case, the phase
$\phi$ consists of a series of kinks of width $l$ the localization
length and located at random positions (with an average spacing
$k_F^{-1} \gg l$).
The low-energy excitations correspond to soliton-like
excitations. In the physical limit we are considering $k_F^{-1} \ll
L_0$, the phase $\phi$ has no kink-like structure but rather smooth
distortions between
random values at a scale of order $L_0$. To get the dynamics, the
approximation we are using only retains the small ``phonon'' like
displacements of the phase $\phi$ relative to the equilibrium position
and no ``soliton'' like excitations.
In the absence of Coulomb interactions the phonon-like excitations
alone, when treated exactly in the classical limit $K\to 0$
are known \cite{vinokur_cdw_exact} to give the
correct frequency dependence of the conductivity
$\omega^2\ln^2(1/\omega)$.
When Coulomb
interactions are included and in the limit where the localization
length is much larger than the interparticle distance, it is not clear
whether soliton-like excitations similar to those considered by Efros
and Shklovskii have to be taken into
account, but in the classical limit $K \to 0$,
phonon modes have a much lower energy than  soliton excitations, and
should dominate the physical behavior of the system.
We would therefore argue that the conductivity is given correctly by
our result (up to possible log corrections) and to behave
in $\omega^2$. If our assumption is
correct the crossover towards the Efros and Shklovskii result when the
disorder becomes stronger would be very interesting to study.

At higher frequencies a crossover occurs above the pinning frequency
$\omega_{\rm pin}$
(typical pinning frequencies are $\omega_{\rm pin} \sim 10^{12} -
10^{14} {\rm Hz}$,
but precise values depend of course on the disorder). Above the
pinning frequency, the WC is not strongly pinned any more but still
feels the scattering on disorder. At the opposite of the LL case, one
now has a universal power-law $\sigma(\omega)\sim 1/\omega^4$. Density
fluctuations give only subdominant corrections \cite{maurey_qwire}.
Above the crossover frequency one recovers the nonuniversal power-law of
the LL  $\sigma(\omega) \sim
(1/\omega)^{4-K_\rho[\omega]-K_\sigma[\omega] - g[\omega]}$
(with scale dependent exponents since one has many impurities).

\begin{figure}
  \centerline{\epsfig{file=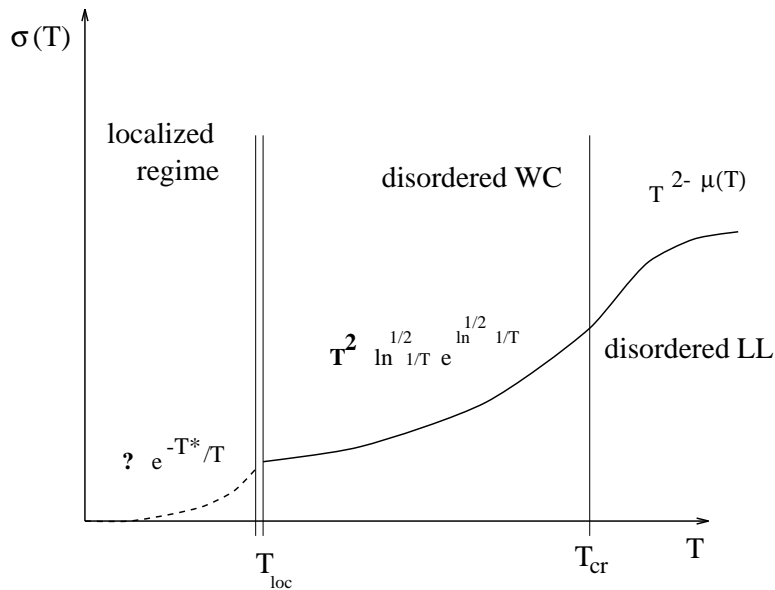,angle=-90,width=10cm}}
   \vspace{0.5cm}
   \caption{Temperature dependence of the conductivity. $T_{\rm
loc}$ and $T_{\rm cr}$ are the boundaries of the localized regime, the
WC regime and the LL regime.}
      \label{condufig}
\end{figure}

The similar three regimes can be found on the temperature dependence of
the conductivity/conductance shown on figure~\ref{condufig}
For the strongly pinned regime
($T < T_{\rm pin}$) one again can naively expect an exponentially
activated  conductivity \cite{maurey_qwire}.
The main characteristic of Coulomb interactions is therefore
to give, for temperatures above the pinning length $T_{\rm pin}$, a
conductivity going
down with decreasing $T$ roughly as $T^2$ (up to the subdominant
corrections). Such a behavior
is characteristic of pinning of the $4 k_F$ density fluctuation i.e.
the WC. Indeed if only
$2 k_F$ scattering was considered, due to the spin fluctuations, even
extremely repulsive charge interactions leading to $K_\rho\to 0$, would
lead to $\sigma(T) \sim T$, i.e. less scattering on the impurities.
Analyzing the temperature dependence of the conductance/conductivity in
one-dimensional wires should therefore provide useful information on the
nature and importance of the interactions, as well
as check the above theories. It is noteworthy that in some
experiments in one dimensional wires, a dependence $G \sim T^2$ was indeed observed
\cite{thornton_wires}.
To probe the correlations in the WC or the LL noise
experiments would prove useful. In a similar
way than for a CDW, such an experiment would provide information on the
periodic nature of the WC, and on its correlation functions.

\section{Conclusions}

We have examined the effects of impurities in a one-dimensional
system. For short-range interactions we have shown that the old results
obtained for long systems with a finite density of impurities and
the more recent results on the conductance of a LL with a single barrier are in
fact identical for weak disorder if one formally lets the effective
scattering go to zero for the macroscopic system. Taking into account
both the distance between impurities and the collective effects leads to
an interesting temperature dependence of the conductivity, that could in
principle be tested in quantum wires.
We also looked at the effects of Coulomb interactions, that lead to the
formation of a WC. Contrarily to the case of the LL, the conductivity
now behaves as $\sigma(T)\sim T^2$, an experimentally testable
prediction.

\noindent
\vspace{1cm}
\begin{center}
\small{\bf Impuret\'e et impuret\'es dans les
syst\`emes unidimensionnels}
\bigskip
\end{center}
\bigskip
{\small Nous examinons les effets du d\'esordre dans les syst\`emes
unidimensionnels.
Nous \'etablissons le lien entre une situation avec un petit nombre
d'impuret\'es, propre \`a d\'ecrire un fil quantique court, et celle
d'un syst\`eme
contenant une densit\'e finie de diffuseurs, i.e. un fil long ou un
syst\`eme macroscopique. Enfin nous \'etudions les effets des
interactions
Coulombiennes sur les propri\'et\'es de transport des fils 1D. Nous
trouvons
une conductivit\'e se comportant en fonction de la temp\'erature comme
$\sigma(T) \sim T^2$.}
\bigskip


\begin{thebibliography}{10}
\bibitem{difference_crossover}
Our results differ from the one of ref. [34]. This difference comes from the
  neglect of the renormalisation of the exponents in [34] and a different, and
  in our opinion, incorrect relation between the conductivity of a macroscopic
  system and the conductance of one impurity instead of equ. (15).
\end{thebibliography}
\end{document}